# Electric Vehicle Adoption Modeling in France: A Systematic Literature Review

K. Widiawati[1], B. M. Sopha[1], N. Rakoto[2]

[1]Department of Mechanical and Industrial Engineering, Universitas Gadjah Mada, Yogyakarta, Indonesia
[2] Department of Control and Industrial Engineering, IMT Atlantique, Nantes, France
(karsiwidiawati@mail.ugm.ac.id)

*Abstract* - **France is one of the pioneer countries in the use of Electric Vehicles (EVs). The French government aims to complete the transition to EVs by 2040. Therefore, modeling related to the adoption of EVs is needed in order to determine the potential policies needed to achieve this goal. This modeling is based on a literature study to identify the factors and the causal relationship between those factors. The systematic literature review (SLR) analysis was performed on 20 journals selected based on PRISMA filtering. From this SLR analysis, five direct factors and four indirect factors were obtained which were then used as the basis for modeling. Based on the model developed, four balancing (B) loops and three reinforcing (R) loops were obtained. From the analysis, it was found that the advertising factor has a goal seeking structure, while the word of mouth, environmentally friendly image, and total cost of ownership factors have an S-shaped structure.**

## I. INTRODUCTION

Nowadays, there is a lot of discussion about global warming caused by greenhouse gas (GHG) emissions. According to Codani et al. [1], European leaders have set goals to reduce $CO_2$ emissions by 40%, improve energy efficiency by 27%, and improve the proportion of renewable energy sources (RES) in the energy mix by 27%. The 27 countries in the European Union (EU) have committed to achieving a share of renewable energy of at least 32% and reducing GHG emissions by at least 40% by 2030, compared to emissions in 1990 [2]. This is in line with the objectives of the Paris Agreement's goals, which aim to limit the rise of the global average temperature to 1.5 °C [3]. These goals can be achieved through energy savings, renewable energy development, and energy efficiency improvement [4]. One of the strategies used by several countries around the world to reduce GHG emissions is the use of electric vehicles (EVs).

The main strategy to reduce GHG emissions from the transportation sector is through the adoption of EVs. EVs are less polluting than those internal combustion engine vehicles (ICEV) because they don't emit any local emissions [5]. Although 67% of electricity is generated from fossil fuels, the use of EVs will result in higher electricity consumption, which will increase the carbon footprint of electricity production [5]. Therefore, the power source has a significant impact on the emissions generated by the use of EVs. According to a study by Nimesh et al. [6], one of the countries identified as ecologically feasible for using EVs is France because this country generates about 72% of its electricity from nuclear power, leading to smaller emissions throughout the electricity generation process.

France is one of the countries that pioneered the use of EVs as road transportation. The French government has proposed a law that would restrict the sale of any new gasoline or diesel vehicles after 2040 and force their replacement with EVs [6]–[8]. This study is a systematic literature review (SLR) that is used to model the adoption of EVs in France with system dynamics to find out the causal and feedback loop relationships involved. This modeling can provide a general overview of the causal structure of the system, which can then be used as a foundation for selecting what policies to implement in order to achieve the government's goal. Simulation models with system dynamics are widely used in studying the effects of policy interventions [9].

This article consists of five sections. In the following section, there is a methodology that contains details of the method. The next section is the results, which contain the SLR results and the model. The fourth section is the discussion, which contains the analysis of the model and potential future work. The last section is a conclusion that contains an overview of the entire contents of the articles.

## II. METHODOLOGY

Systematic literature review (SLR) and system dynamics are the two methodologies used in this study. In order to determine what influence the adoption of EVs in France, a SLR was carried out using the search string (("electric vehicle*") AND ("adoption" OR "buy*") AND ("france" OR "french")). The search was limited to the types of articles in Scopus, Ebsco, and ProQuest. After filtering the 43 articles that were found during the search, there are 20 articles that were to be analyzed remained. The PRISMA diagram in Figure 1 provides information on the performed filtering procedure.

The system dynamics modeling carried out in this article is limited to qualitative modeling in the form of a causal loop diagram (CLD), which illustrates the causal relationship between the factors involved in the EV adoption system in France. The use of system dynamics was chosen because this method applies a system thinking approach to improve the understanding of complex systems that have feedback interactions among the variables in the system [9]. A previous systematic literature study was used to determine the relationship among these factors.

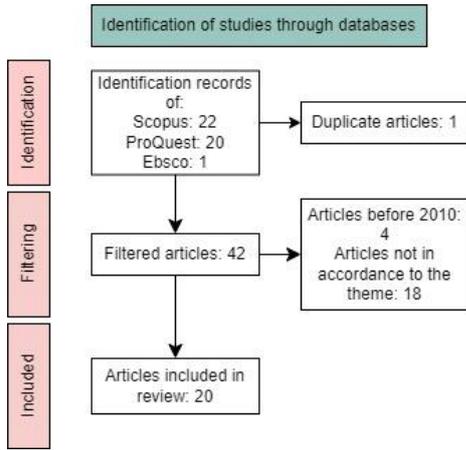

Fig. 1. The procedure for selecting articles using PRISMA.

## III. RESULTS

### A. Bibliometric analysis

The bibliometric analysis conducted in this article is used to determine the trend of research topics from the selected articles. The bibliometric analysis was conducted with VOSviewer and can be seen in Figure 2.

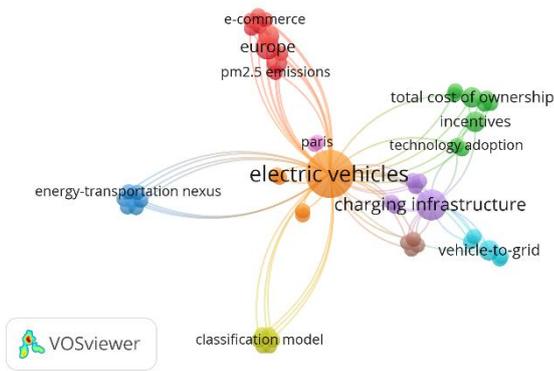

Fig. 2. Bibliometric analysis using VOSviewer.

Based on the bibliometric analysis results, there are 7 clusters shown in orange, purple, blue, yellow, dark blue, red, and green. The orange cluster is related to EVs, the purple cluster is related to EV charging infrastructure, the blue cluster is related to the vehicle to grid, the yellow cluster is related to the EV technology, the dark blue cluster is related to energy transportation, the red cluster is related to European targets, and the green cluster is related to the total cost of ownership. The results of this bibliometric analysis provide an early overview of the information contained in the selected articles before conducting further analysis.

### B. Profile analysis

Profile analysis in this article refers to the origin of the first author institution in every article selected. This profile analysis is used to determine the author's origin trend of the selected article. The results of the profile analysis can be seen in Figure 3.

The results of the analysis show that 80% of the first author of the selected articles were from Europe. This could be due to the fact that the study focus of this article is limited to France cases, so most researchers who conduct research with a case study in this country are mostly from European countries. With this mostly European researcher profile, it is expected that the results of the analysis obtained can be more relevant to the situation in France.

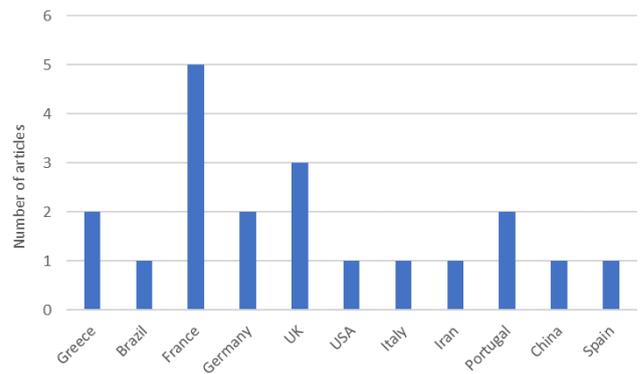

Fig. 3. Profile of the article's first author.

### C. EV adoption influential factors in France

Factors that affect EV demand in France can be divided into direct and indirect factors. The direct factors are factors that directly influence the demand for EVs, while the indirect factors are factors that indirectly influence, usually influencing the direct factors. Based on the analysis conducted on 20 articles obtained from filtering with PRISMA, the following table presents these factors in more detail, along with their sources.

TABLE I
INFLUENTIAL FACTORS OF EV DEMAND IN FRANCE

| Direct | Indirect |
|---|---|
| Technical readiness [10]–[13], [16]–[18], [25]–[27] | Government subsidy [10], [14], [15], [27], [28] |
| The difference in total cost ownership [8], [10], [11], [14], [15], [17], [18], [28]–[30] | Taxes [8], [10], [14], [15], [27] |
| Advertising effects [7], [8], [10] | Green energy ratio [8], [10], [17], [19], [20] |
| Word of mouth effects [7], [10] | Policy intervention [12], [18], [27], [29] |
| Environmentally friendly image [11], [21] | |

The technical readiness factors mentioned in Table 1 include EV features, charging infrastructure, and battery switch infrastructure. The EV features in this factor are the number of models available on the market [10] and the range of vehicles [11]. The charging infrastructure includes regular charging stations [8] as well as the development of fast charging technology [12]. The use of battery switch stations reduces the need for distribution network electricity for charging EVs and only needs to provide battery switch stations. In addition, using a battery switch will make it easier to collect batteries because batteries are anonymous and will definitely be collected there. The use of a battery switch will also help to ensure ideal charging circumstances, which will cause the battery to be preserved and have a longer shelf life [13].

The difference in the total cost of ownership factor mentioned in Table 1 relates to the comparison between the total cost of EVs and ICEVs. According to Flaris et al. [14], a significant reason for the slow adoption of EVs in society is the high purchase price of EVs. Lévay et al. [15] state that the total cost of ownership of an EV consists of the purchase price of the vehicle, taxes, subsidies, electricity prices, and the resale value of the vehicle. In France, the total cost of ownership of small EVs is higher than the total cost of ownership of small ICEVs but lower for large vehicle types [14]. According to Lévay et al. [15], small EV demand is more price-responsive than large EV demand, so subsidizing small EVs would significantly increase EV demand.

Potential adopters for EVs in France are quite large [8]. This shows good potential for the future diffusion of EVs. The development of fast charging technology has led to the development of EV car sharing. Car sharing is useful for the efficient use of EVs and also to promote EVs by providing the experience of using EVs. This will reduce the assumption stating that EVs are not technically mature and introduce a good performance both in terms of the EV product itself and supporting infrastructure such as battery charging and switch stations [16].

The main electrical energy source in France is nuclear. According to Doufene et al. [17] and Fuinhas et al. [18], the use of nuclear power as the basis of electricity sources leads to lower electricity prices, making the use of EVs relatively more profitable compared to many other countries in the world. The use of EVs and the transition to renewable energy will help achieve long-term climate neutrality by 2050 [19]. EV adoption requires effective political planning to achieve net zero emissions [18]. This can be found in France's vehicle decarbonization plan that has been made since 2009 [15], and France's plan to ensure no new gasoline or diesel cars will be sold after 2040 [7], [8].

Consumable batteries available from EV applications can meet the stationary energy needs of PV systems [20]. In France, the batteries available for a second life could cover 26-87% of the stationary storage needs of 42 GW PV installations by 2030 [20]. The use of EVs in line will support the development of renewable energy generation. The utilization of EV batteries as energy storage will help reduce investment costs in the development of renewable energy power plants [21]. With the increase in renewable energy, it will increase the green charging ratio and improve the environmentally friendly image of EVs. Based on a study from Ortar et al. [11], one of the factors hindering the diffusion of EVs is uncertainty regarding the benefits of using EVs to reduce emission levels. Therefore, an environmentally friendly image of EVs is needed.

### D. Causal loop diagram (CLD) of EV adoption in France

The modeling conducted in this article only focuses on qualitative modeling using CLD. This CLD is intended to describe the causal relationship that occurs in the system. The details of the causal relationships that occur in the EV adoption system in France can be seen in Figure 4.

The modeling method is based on an SLR that has been conducted. There are 7 loops that were formed from the model that had been made. In general, there are 2 types of loops that can be formed in the system dynamics model, i.e. balancing (B) and reinforcing (R). The loop formed then will determine the structure and behavior of the system. The details of the loop formed are described below:

**Loop 1: B1 Advertising**
Demand – EV sales – Adoption rate – EV potential adopters – Adoption from advertising

**Loop 2: B2 Word of mouth**
Demand – EV sales – Adoption rate – EV potential adopters – Contact rate of potential adopters and adopters – Adoption from word of mouth

**Loop 3: R3 Word of mouth**
Demand – EV sales – Adoption rate – EV adopters – Probability of contact – Contact rate of potential adopters and adopters – Adoption from word of mouth

**Loop 4: R4 Environmentally friendly image**
Demand – EV sales – Battery electric vehicle – PV based energy – Green charging ratio – Environmentally friendly image

**Loop 5: B5 Environmentally friendly image**
Demand – EV sales – Battery electric vehicle – PV based energy – Energy generation – Green charging ratio – Environmentally friendly image

**Loop 6: R6 Total cost**
Demand – EV sales – Battery electric vehicle – PV based energy – Green charging ratio – Electricity price – EV total cost – EV total cost-ICEV total cost

**Loop 7: B7 Total cost**
Demand – EV sales – Battery electric vehicle – PV based energy – Energy generation – Green charging ratio – Electricity price – EV total cost – EV total cost-ICEV total cost

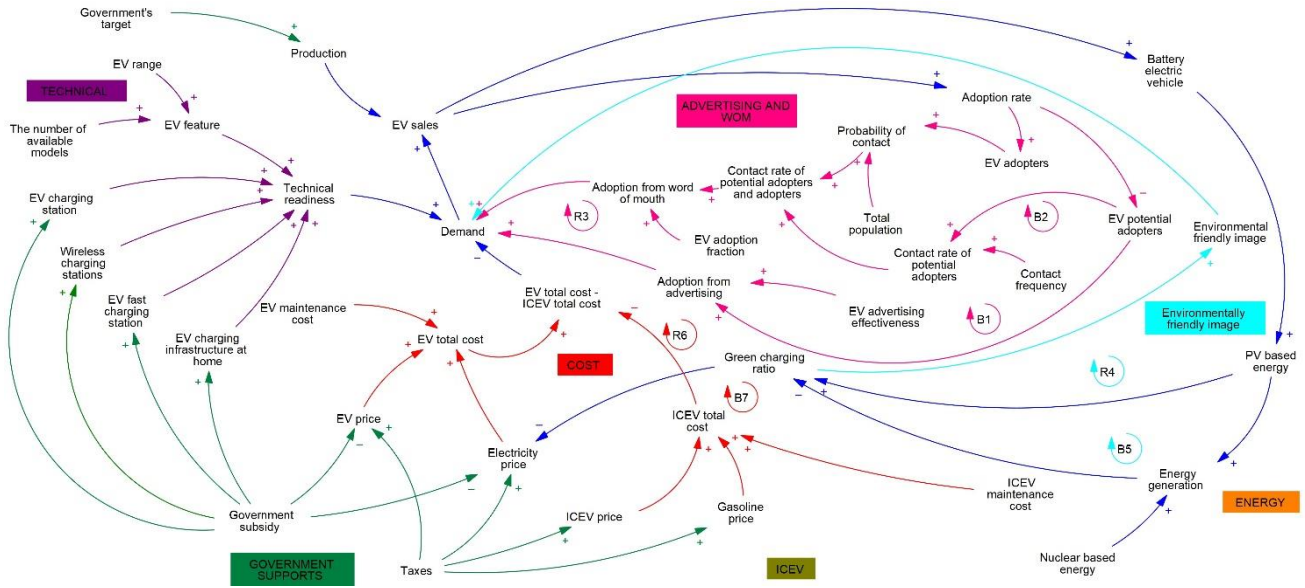

Fig. 4. CLD of EV adoption in France.

## IV. DISCUSSION

### A. Potential policy design

The results of qualitative analysis using CLD show the main structure of several factors. In the advertising factor, it is found that the trend value to be obtained is goal seeking. Based on Sterman [22], the structure of a system consisting of a balancing loop is goal seeking, where the increase or decrease of the value will stop at a certain point. The efforts that can improve and accelerate its growth include increasing the number of potential adopters both by improving their understanding of the benefits of using EVs and by improving public awareness of issues related to environmental pollution.

The word of mouth factor obtained one B loop and one R loop which, based on Sterman [22], will form an S-shaped growth structure. This also happens to the environmentally friendly image and total cost factors. These two factors also have an S-shaped growth structure.

In terms of the word of mouth factor, increasing the adoption of EVs can be achieved by increasing contact between adopters and potential adopters. This can be done by implementing car sharing. The market for EVs in the long term cannot be achieved without the support of social commitment [23].

In terms of the environmentally friendly image factor, actions that can be implemented to increase EV adoption include increasing the green charging ratio through increasing the use of renewable energy, one of which is the development of PV energy generation.

In terms of the total cost factor, possible ways to increase EV adoption are by reducing the total cost of EV ownership through subsidies, tax exemptions, and reducing electricity prices. In addition, it is necessary to limit the use of ICEV by establishing regulations, increasing fuel prices, and implementing emission taxes.

Besides the factors above, based on the model, the technical readiness factor is also strongly influencing the adoption of EVs, either related to charging infrastructure or related to the performance of EVs. The influence of the performance of EVs is in line with the results of the previous study by Widiawati et al. [24] who found that public acceptance of EVs is significantly influenced by the performance of the EV.

### B. Future Work

This article only focuses on qualitative modeling of EV adoption in France using CLD. The future research that will be conducted is continuing this qualitative modeling by quantitatively computing using stock and flow diagrams (SFD) to get a more specific result that can be simulated for real systems.

## V. CONCLUSION

This study is intended to qualitatively model the adoption of EVs using CLD to understand the causal structure in the system and determine potential policies to assist in achieving the desired targets of the France government in the future. As a result of the SLR that has been conducted, there are five direct factors and four indirect factors that influence the adoption of EVs. Based on the model developed, it is found that the structure of the model for the advertising factor is goal seeking, whereas for the word of mouth, environmentally friendly image, and total cost factors, the structure are S-shaped.

ACKNOWLEDGMENT

This research is an outcome of the PhD Program Mobility scheme at IMT Atlantique, France for the track Management and Optimization of Supply Chains and Transport in the frame of the FICEM consortium, with the Erasmus+ Scholarship.